# End-to-End Trainable Self-Attentive Shallow Network for Text-Independent Speaker Verification

*Hyeonmook Park*[1], *Jungbae Park*[1,2] *, *Sang Wan Lee*[1,2,3]*

[1]Humelo Inc.
[2]Korea Advanced Institute of Science and Technology (KAIST)
[3]KAIST Institute for Artificial Intelligence
*Corresponding authors

`mook1993@dankook.ac.kr, jbp0614@kaist.ac.kr, sangwan@kaist.ac.kr`

## Abstract

Generalized end-to-end (GE2E) [1] model is widely used in speaker verification (SV) fields due to its expandability and generality regardless of specific languages. However, the long-short term memory (LSTM) [2] based on GE2E has two limitations: First, the embedding of GE2E suffers from vanishing gradient, which leads to performance degradation for very long input sequences. Secondly, utterances are not represented as a properly fixed dimensional vector. In this paper, to overcome issues mentioned above, we propose a novel framework for SV, end-to-end trainable self-attentive shallow network (SASN), incorporating a time-delay neural network (TDNN) and a self-attentive pooling mechanism [3, 4] based on the self-attentive x-vector system [5, 6] during an utterance embedding phase. We demonstrate that the proposed model is highly efficient, and provides more accurate speaker verification than GE2E. For VCTK dataset [7], with just less than half the size of GE2E, the proposed model showed significant performance improvement over GE2E of about 63%, 67%, and 85% in EER (Equal error rate), DCF (Detection cost function), and AUC (Area under the curve), respectively. Notably, when the input length becomes longer, the DCF score improvement of the proposed model is about 17 times greater than that of GE2E.

## 1. Introduction

Speaker verification (SV) refers to the task of verifying the identity of a person based on the characteristics of the voice. There are two different approaches to SV: text-dependent (TD) and text-independent (TI). In TD-SV, all stages of the SV process rely on a predefined text. On the other hand, TI-SV does not impose the constraints of the spoken phrases by the speaker, making it much more demanding than the TD-SV. In general, the SV process consists of three steps: development, enrollment, and evaluation. In the development step, a background model for speaker expression is created. The enrollment step uses the background model to generate a speaker model for a new user. Finally, the evaluation phase confirms or rejects the identity of the test utterance by comparing expression with previously generated speaker models.

One of the powerful approaches in SV applications was i-vector based systems [8]. It is based on the combination of an encoder module that generates speaker representation and a discriminative module that compares speaker embedding vectors of the encoder. The i-vector system requires independent training of the encoder and the discriminative modules, essentially increasing the size of the complexity of the system. Recent studies have attempted to design SV systems in a way to be trainable in an end-to-end manner [9, 10, 11, 12, 13]. In this approach, the speaker embedding and the discrimination part are have been integrated into a single network, such that it requires minimal configuration and assumptions. This single network approach improves test accuracy by precluding accumulation of errors in each independent part. Especially, GE2E [1] model has wide applicability. For example, GE2E achieves good performance on both TD-SV and TI-SV tasks, as well as on speaker diarization [14, 15], multi speaker TTS [16], speech-to-speech translation [17] and voice filter [18].

However, GE2E has a few limitations. Firstly, GE2E has a LSTM based embedding structure, making it difficult to process long inputs. This can be ascribed to the vanishing gradient [19], which often occurs when processing long input sequences. According to the study [20] examining the effect of input length on SV accuracy, the longer length of the input speech can help the model improve the SV accuracy. Secondly, the encoder of GE2E considers all frames equally, failing to accommodate the fact that each frame conveys a different amount and type of information for speaker identification. For example, some frames may have valuable information to identify speakers, whereas some have little information. This means that model's computational power is also consumed to learn the frames with unnecessary information, which may have bad influence on computational efficiency and accuracy.

To cope with these issues, we propose a novel framework called self-attentive shallow network (SASN). SASN has the following advantages. First, its embedding module is based on a combination of a TDNN and a self-attention mechanism. TDNN's sub-sampling method makes it possible to identify discriminative patterns in a wide temporal context while reducing computational load. In addition, its parallel sampling operation makes training faster than RNNs whose outputs are autoregressively calculated. And its pooling layer, based on a self-attention mechanism allowing the model to identify an important time window to focus on [21], improves embedding efficiency. Second, a single end-to-end architecture of SASN not only prevents accumulation of errors between different independent components, but also makes it easier to configure the hyperparameter setting.

Finally, we found that the efficiency and performance of SASN can be further improved by introducing a batch-wise loss, as done in [1], because it allows the model considers various relationships between multiple data at each step of the training process. As a result, when tests were carried out on with VCTK [7] datasets, SASN not only shows 63%, 67% and 85% performance improvement over the baseline GE2E in terms of EER, DCF and AUC, respectively, and notably, our model size is only 58% of GE2E. Moreover, we found that for longer inputs, thanks to the self-attentive components of SASN, performance improvement was remarkable. When the input length was increased from 180 to 300 frames, the performance improvement in terms of DCF was 17 times greater than that of GE2E. Full details on performance comparison are provided in Section 5.

## 2. Related works

### 2.1. Self-Attentive Speaker Embedding SV

The process of self-attentive x-vector system [6] consists of two phases: embedding and scoring. In the embedding process, the speaker discriminative TDNN is trained to produce speaker-specific representation called x-vectors. In the scoring process, a probabilistic linear discriminant analysis (PLDA) [22] backend is used to compare pairs of x-vectors. In the embedding stage, each frame of the utterance embedding resulting from previous TDNN layers represents context information. Then the self-attentive pooling layer aggregates over frame-level vectors by deriving a mean and a standard deviation from the outputs of the final layer of TDNN. One challenge of this approach is that each of the embedding and scoring module should be trained independently, making it difficult to find the optimal configuration of train/test environments and datasets. Besides, independently trained modules can potentially accumulate errors.

To resolve this issue, we integrate the embedding part and the scoring part into a single network that is trainable in a end-to-end fashion. As a result, the total number of parameters of the whole system of SASN is 1940K, which is only less than half the size of the embedding part of the x-vector model (4200K)[5].

### 2.2. Generalized End-to-End loss For Speaker Verification

GE2E [1] model maps the utterance to an embedding vector by using LSTM and scores similarity between them. Each part has an independent role and is optimized jointly. The loss function of GE2E was designed for batch training; it can learn a large number of utterances at once. Each utterance embedding is used to obtain the corresponding speaker embedding. Subsequently, the similarity matrix was calculated based on the cosine similarity between each utterance and each speaker. The cost function is optimized in a way that the similarity between each utterance and the corresponding speaker is maximized.

We introduced TDNN and the attention pooling layer to overcome the limitation of LSTM-based embedding methods, as mentioned in Section 1. Owing to TDNN, along with the attentional pooling layer, the performance of the model is significantly increased without any temporal information loss. The results of the performance analysis are summarized in Table 2 and Figure 2.

## 3. Proposed framework

Our SASN model consists of two modules. The encoder generates the representation vector from each utterance feature. The scoring module then compares the embedding vectors of the encoder and scores them. These two parts are jointly trained within a single network in a fully end-to-end fashion, as shown in Figure 1.

### 3.1. Context-aware, self-attentive encoder module

The purpose of the encoder is to produce an utterance embedding. It is based on the encoder architecture introduced in the x-vector baseline system [8, 9]. The frame-level input, a 40-dimensional feature vector, passes through the TDNN that consists of the three layers from l1 to l3. The layer l3 thus processes contextual temporal information. The configuration of TDNN is outlined in Table 1.

Table 1. *Configuration of TDNN Architecture*

| Layer | Context window | # Context | In×Out |
|---|---|---|---|
| l1 | $(t-2, t+2)$ | 5 | $200 \times 512$ |
| l2 | $\{t-2, t, t+2\}$ | 9 | $1536 \times 512$ |
| l3 | $\{t-3, t, t+3\}$ | 15 | $1535 \times 512$ |

The output vector of TDNN is passed to the attention layer. The structure of the attention layer is shown in Figure 1. Suppose that an utterance segment of length $T$ is represented as the sequence of vectors $H = \{h_1, h_2 \ldots h_T\}$ and the number of attention heads is $d_r$. In the first attention layer, $H$ of the size $512 \times T$ is compressed into the fixed-length $E_{single}$ of the size $512 \times d_r$.

$$A_{single} = softmax(f(H^T W_1)W_2), \quad (1)$$

$$E_{single} = \|HA_{single}\|_2, \quad (2)$$

where the size of each term is as follows:
$W1: 512 \times d_a$; $W2: d_a \times d_r$; $A_{single}: T \times d_r$;
The $f(\cdot)$ is a ReLUs activation function. The $softmax(\cdot)$ is computed column-wise. The first attention layer multiplies each $h_t$ by its weight differentially according to the amount of information necessary for speaker identification. As a result, each column of $E_{single}$ encodes different utterance characteristics as follows:

$$\begin{array}{c} Col_c = \alpha_1 h_1 + \alpha_2 h_2 + \cdots + \alpha_T h_T, \\ (c: column\ index,\ \sum_{i=1} \alpha_i = 1) \end{array} \quad (3)$$

Furthermore, we also found that stacked self-attention enhances performance, consistent with previous demonstrations [23, 24, 25]. In SASN, an additional attention operation can be performed as follows:

$$A_{double} = softmax\left(g(E_{single}^T W_3)\right), \quad (4)$$

$$E_{double} = \|E_{single} * g(A_{double}^T)\|_2, \quad (5)$$

The size of $W3$ is $512 \times 1$, and therefore the size of $A_{double}^T$ is $1 \times d_r$. The function $g(\cdot)$ is a tile operation that is repeatedly applied 512 times based on the row axis. The optional second attention layer performs column-wise multiplication on $E_{single}$. The output $E$ ($E_{single}$ or $E_{double}$) through the attention layer enters the pooling layer. Finally, the 1-dimensional speaker embedding vector $e$ is concatenated with a mean and standard deviation of $E$ on a time axis.

### 3.2. Scoring module

The scoring module is intended to quantify the similarity between utterances embedding with speaker embedding. The input batch consists of the multiple utterance embedding vectors forwarded from the encoder module. The input batch is $N \times M$ matrix, where the utterances are from $N$ different speakers and each speaker has $M$ utterances. Each embedding vector is given by $x_{ji}$ ($1 \leq j \leq N$, $1 \leq i \leq M$) for the *j-th* speaker and the *i-th* utterance. The centroid of the embedding vectors $c_k$ from the $k$-$th$ speaker is obtained from the average value of the utterance embedding $[e_{k1}, e_{k2} ... e_{kM}]$. However, when comparing the centroid of $k$-$th$ speaker with utterance embedding $e_{ki}$ of $k$-$th$ speaker, the $k$-$th$ speaker centroid is replaced with the average value excluding $e_{ki}$ for stable training and avoiding trivial solution. The similarity matrix $S$ quantifies similarity between all embedding vectors $e_{ji}$ and centroid $c_k$ pairs as follows:

$$S_{ji,k} = w \cdot \cos(e_{ji}, c_k) + b \quad (w > 0), \quad (6)$$

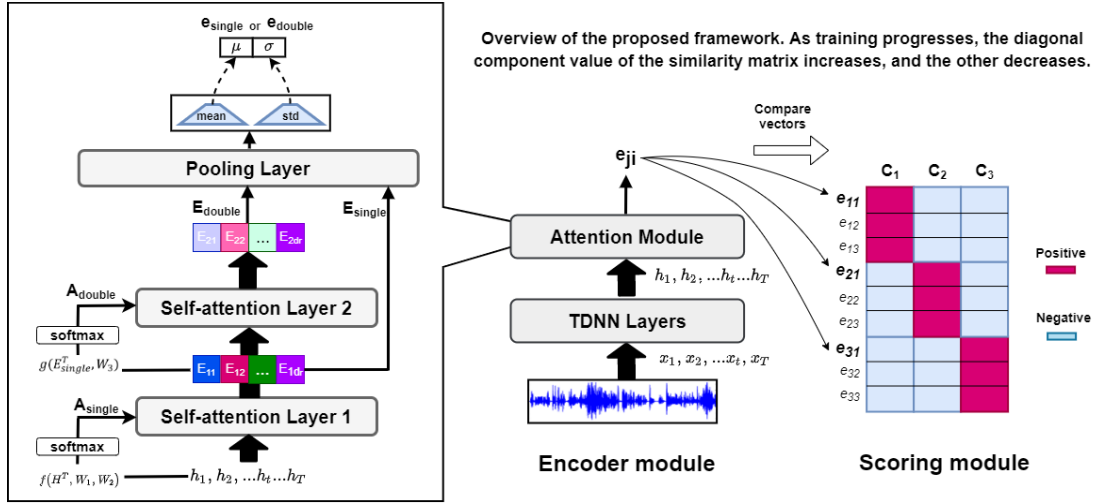

Figure 1. Structure of self-attention module and system overview. The input entering the pooling layer is the $E_{single}$ or $E_{double}$

The loss function, shown in below (7), makes the similarity between utterance embedding and the corresponding speaker embedding 1, whereas the similarity between different speaker embedding is set to -1. The penalty term $P$ described in [13] is intended to reduce redundancy errors in the attention matrix $A_{single}$.

$$L(e_{ji}) = -S_{ji,j} + \log \sum_{k=1}^{N} \exp(S_{ji,k}) + \alpha P, \quad (7)$$

In the SV model, the loss function (7) is used to jointly train the encoder and the scoring component

## 4. Experiments

### 4.1. Simulation setting

Our simulation setting mostly follows what the baseline system used [1, 9]. In the feature extraction process, raw audio signals are transformed into the vector of the 25ms frame with 10ms step size. Then we compute 40-dimension log-mel-filter bank energies for each frame. In the encoder module, the input frame length is more than 180 frames (180, 300, 600). The network was trained with stochastic gradient descent (SGD) using the learning rate 0.01; clipping gradient descent was not applied. We set the attention head number $d_r$ to 5, 10 and 20 respectively.

### 4.2. Datasets

We used the following datasets to train and evaluate. First, the VCTK [7] dataset is an English dataset consist of recorded voice of 109 speakers. Each speaker reads about 400 sentences of different news articles. It contains 13100 voices in total (44 hours). Second, we used TIMIT [26] dataset, which contains a total of 6300 voices (5.4 hours) consisting of 10 sentences recorded by each of the 630 speakers in the eight major dialect regions of the United States. The third dataset used is 20SER; it is a Korean dataset with some artifact noise due to it being recorded using condenser microphone. This dataset consists of 5593 voices with 20 speakers recorded while reading the drama script (4.7 hours). The dataset not only varies the number of sentences read by each speaker from 9 to 400, but the length of the sentence also varies from 0.89 sec to 17.02 sec and the average length is 3.03 sec.

### 4.3. Evaluation

We ran experiments in three different scenarios. In the first case (Case1 of Table2), the VCTK dataset is used. The train dataset consists of 90 speakers, and the test dataset consists of the remaining 19 speakers. In the second case (Case2 of Table2), the 20SER dataset is used. The train was carried out with 16 speakers' data, and the evaluation was with the remaining 4 speakers. In the third case (Case3 of Table2), all the above datasets were used; VCTK, and TIMIT datasets were used for training, while 20ser datasets for evaluation. Note that in all cases, the male and female ratios were almost identical both for the training and the test dataset.

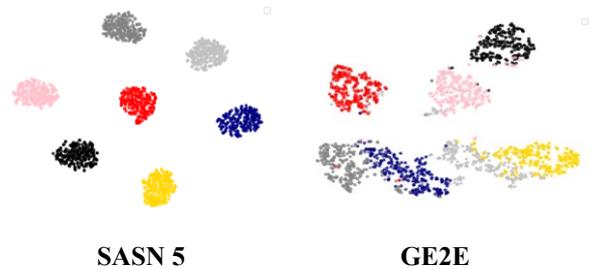

Figure 2. *Visualization with t-SNE of the latent space of each model on 20SER dataset*

Table 2. *Performance comparison between SASN and GE2E (test accuracy)*

|         | Case1 | | | Case2 | | | Case3 | | |
|         | Train: VCTK   Test: VCTK | | | Train: 20SER   Test: 20SER | | | Train: VCTK+TIMIT   Test:20SER | | |
|         | EER(%) | DCF(%) | AUC | EER(%) | DCF(%) | AUC | EER(%) | DCF(%) | AUC |
|---|---|---|---|---|---|---|---|---|---|
| GE2E    | 6.05 | 0.73 | 98.40 | 18.95 | 0.99 | 88.43 | 25.72 | 1.04 | 80.51 |
| SASN 5  | 2.76 | 0.3  | 99.63 | **9.32** | 0.79 | **96.43** | 17.28 | 0.75 | **88.89** |
| SASN 10 | 2.49 | 0.25 | 99.73 | 9.62 | 0.77 | 96.10 | **17.02** | 0.75 | 88.82 |
| SASN 20 | **2.23** | **0.24** | **99.76** | 9.53 | **0.76** | 96.19 | 17.14 | **0.74** | 88.83 |

## 5. Simulation results

The performance comparison between SASN and GE2E for each experiment case is shown in Table 2. Note that the model using only the attention layer 1 and layer 2 is called Single SASN and Double SASN, respectively. Also, the number at the end of the SASN notation indicates the number of attention heads $d_r$ (5, 10, 20); e.g., SASN 5 means the model with 5 attention heads. We used three different performance measures: EER, DCF, and AUC. The formula of each metric is given as follows.

$$EER = (FAR + FRR)/2,$$
$$(threshold, \theta = argmin(|FAR - FRR|)) \quad (8)$$

$$DCF = P_{target}FRR + (1 - P_{target})FAR, \quad (9)$$
$$(P_{target} = 0.01)$$

$$AUC = \sum_{\theta=-1}^{1} TRR_\theta (TPR_{\theta-1} - TPR_\theta), \quad (10)$$
$$(threshod, \theta)$$

Our model significantly outperformed GE2E in terms of all of the three types of performance metric. First of all, the SASN 5 used 58% less resources in terms of model size than the GE2E. Intriguingly, the SASN achieves better accuracy with only 1,940K parameters than the GE2E which has 4,654K parameters. In Case1 (English dataset VCTK), the EER score of the SASN was 63% higher than GE2E. The DCF score was 67% higher, and the AUC score was 85% higher than its counterpart. In Case2 (Korean Dataset 20SER), the SASN showed better performance up to 51%, 23%, and 69% in EER, DCF, and AUC respectively. In Case3 in which all the datasets were used, the performance improvement of the SASN was 34%, 29%, and 43% in terms of EER, DCF, and AUC, respectively. The result of Case3 experiment shows that even lightweight SASN is also able to successfully learn the cross-language task.

Table 3. *Single SASN vs Double SASN on the TIMIT dataset*

|  | softmax (%) | DCF(%) | AUC |
|---|---|---|---|
| Double SASN | **2.00** | .684 | **98.77** |
| Single SASN | 2.47 | **.682** | 98.55 |
| GE2E | 5.57 | .768 | 97.93 |

The 2-dimensional visualization with t-SNE (Stochastic Neighbor Embedding) of the learned latent space of the baseline

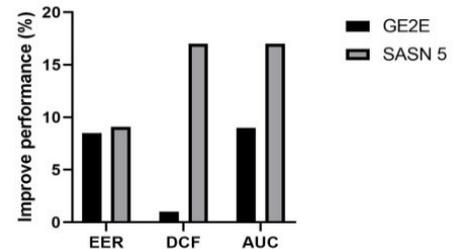

Figure 3. *Accuracy improvement when the input length is increased from the 180 to 300 frames*

GE2E and SASN 5 used in the Case3 is shown in Figure 2. Each point in Figure3 is color-coded according to the speaker ID. It is clear that the embedding quality of the SASN 5 model is better than that of the GE2E. Table 3 shows that the difference in accuracy between the single and double attention mode of SASN for the TIMIT dataset. The double mode is 19% and 15% higher than the single attention mode in terms of the softmax and AUC score, respectively. Finally, to roughly assess performance improvement of the SV model as a function of the input utterance length, we re-ran the Case2 experiment with the increased input length from 180 to 300 frames. As shown in Figure 3, the DCF performance gain of the SASN 5 for the long input was 17 times larger than that of the GE2E.

## 6. Conclusion

This paper aims to overcome the limitations of the GE2E [1], such as reduced efficiency for lengthy inputs and a lack of efficient resource allocation to different frames during embedding. To resolve these issues, we propose a new SV framework. It is inspired by the complementary speaker embedding method from the self-attentive x-vector system [6]. Compared to the GE2E, the proposed model showed 63%, 67%, and 85% performance improvement in EER, DCF, and AUC score, even though the model size is reduced by 58%. In a subsequent analysis, when the input length is increased from 1.8sec to 3sec, we found that the improvement was about 17 times greater than that of the GE2E in terms of DCF.

## 7. Acknowledgements

This work was partly supported by    SeoulR&BD Program (CY190019) and by Institute of Information & Communications Technology Planning & Evaluation (IITP) grant funded by the Korea government (MSIT) (No.2019-0-01371, Development of brain-inspired AI with human-like intelligence)

# 8. References

[1] Wan, Li, et al. "Generalized end-to-end loss for speaker verification." *2018 IEEE International Conference on Acoustics, Speech and Signal Processing (ICASSP)*. IEEE, 2018.

[2] Hochreiter, Sepp, and Jürgen Schmidhuber. "Long short-term memory." *Neural computation 9.8* (1997): 1735-1780.

[3] Peddinti, Vijayaditya, Daniel Povey, and Sanjeev Khudanpur. "A time delay neural network architecture for efficient modeling of long temporal contexts." *Sixteenth Annual Conference of the International Speech Communication Association. 2015.*

[4] Garcia-Romero, Daniel, and Alan McCree. "Stacked Long-Term TDNN for Spoken Language Recognition." *INTERSPEECH*. 2016.

[5] D. Snyder, D. Garcia-Romero, G. Sell, D. Povey, and S. Khudanpur, "X-vectors: Robust DNN embeddings for speaker recognition," in Proceedings of the *IEEE International Conference on Acoustics, Speech, and Signal Processing*, 2018.

[6] Zhu, Yingke, et al. "Self-Attentive Speaker Embeddings for Text-Independent Speaker Verification." *Interspeech*. 2018.

[7] C. Veaux, J. Yamagishi, and K. MacDonald. CSTR VCTK corpus: English multi-speaker corpus for CSTR voice cloning toolkit, 2012. URL http://dx.doi.org/10.7488/ds/1994.

[8] Najim Dehak, Patrick J Kenny, Reda Dehak, Pierre Du-mouchel, and Pierre Ouellet, "Front-end factor analysis for speaker verification," *IEEE Transactions on Audio, Speech, and Language Processing*, vol. 19, no. 4, pp. 788–798, 2011.

[9] Ehsan Variani, Xin Lei, Erik McDermott, Ignacio Lopez Moreno, and Javier Gonzalez-Dominguez, "Deep neural networks for small footprint text-dependent speaker verification," in *Acoustics, Speech and Signal Processing (ICASSP)*, 2014 IEEE International Conference on. IEEE, 2014, pp. 4052–4056.

[10] Yu-hsin Chen, Ignacio Lopez-Moreno, Tara N Sainath, Mirko Visontai, Raziel Alvarez, and Carolina Parada, "Locallyconnected and convolutional neural networks for small footprint speaker recognition," in Sixteenth Annual Conference of the *International Speech Communication Association*, 2015.

[11] Chao Li, Xiaokong Ma, Bing Jiang, Xiangang Li, Xuewei Zhang, Xiao Liu, Ying Cao, Ajay Kannan, and Zhenyao Zhu, "Deep speaker: an end-to-end neural speaker embedding system," *CoRR*, vol. abs/1705.02304, 2017.

[12] Shi-Xiong Zhang, Zhuo Chen, Yong Zhao, Jinyu Li, and Yifan Gong, "End-to-end attention based text-dependent speaker verification," *CoRR*, vol. abs/1701.00562, 2017.

[13] Seyed Omid Sadjadi, Sriram Ganapathy, and Jason W. Pelecanos, "The IBM 2016 speaker recognition system," *CoRR*, vol. abs/1602.07291, 2016.

[14] Q. Wang, C. Downey, L. Wan, P. A. Mansfield, and I. L. Moreno, "Speaker diarization with lstm," in *International Conference on Acoustics, Speech and Signal Processing (ICASSP)*. IEEE, 2018, pp. 5239–5243.

[15] A. Zhang, Q. Wang, Z. Zhu, J. Paisley, and C. Wang, "Fully supervised speaker diarization," *arXiv preprint arXiv:1810.04719*, 2018.

[16] Y. Jia, Y. Zhang, R. J. Weiss, Q. Wang, J. Shen, F. Ren, Z. Chen, P. Nguyen, R. Pang, I. L. Moreno et al., "Transfer learning from speaker verification to multispeaker text-to-speech synthesis," in *Conference on Neural Information Processing Systems (NIPS)*, 2018.

[17] Y. Jia, R. J. Weiss, F. Biadsy, W. Macherey, M. Johnson, Z. Chen, and Y. Wu, "Direct speech-to-speech translation with a sequenceto-sequence model," *arXiv preprint arXiv:1904.06037*, 2019.

[18] Wang, Quan, et al. "Voicefilter: Targeted voice separation by speaker-conditioned spectrogram masking." *arXiv preprint arXiv:1810.04826* (2018).

[19] Hochreiter, Sepp. "The vanishing gradient problem during learning recurrent neural nets and problem solutions." *International Journal of Uncertainty, Fuzziness and Knowledge-Based Systems* 6.02 (1998): 107-116.

[20] Pelecanos, Jason, Upendra Chaudhari, and Ganesh Ramaswamy. "Compensation of utterance length for speaker verification." *ODYSSEY04-The Speaker and Language Recognition Workshop*. 2004.

[21] Lin, Zhouhan, et al. "A structured self-attentive sentence embedding." *arXiv preprint arXiv:1703.03130* (2017).

[22] S. Ioffe, "Probabilistic linear discriminant analysis," in *European Conference on Computer Vision*. Springer, 2006, pp. 531–542.

[23] Afouras, Triantafyllos, et al. "Deep audio-visual speech recognition." *IEEE transactions on pattern analysis and machine intelligence* (2018).

[24] Devlin, Jacob, et al. "Bert: Pre-training of deep bidirectional transformers for language understanding." *arXiv preprint arXiv:1810.04805* (2018).

[25] Peters, Matthew E., et al. "Deep contextualized word representations." *arXiv preprint arXiv:1802.05365* (2018).

[26] J. S. Garofolo, L. F. Lamel, W. M. Fisher, J. G. Fiscus, and D. S. Pallett, "DARPA TIMIT acoustic-phonetic continous speech corpus CD-ROM. NIST speech disc 1-1.1," *NASA STI/Recon technical report*, vol. 93, 1993. URL https://catalog.ldc.upenn.edu/LDC93S1